# Quantum limit in subnanometre-gap tip-enhanced nanoimaging of few-layer MoS$_2$

Yingchao Zhang[1,2*], Dmitri V. Voronine[1,3*#], Shangran Qiu[1,2], Alexander M. Sinyukov[1], Mary Hamilton[3], Alexei V. Sokolov[1,3], Zhenrong Zhang[3] and Marlan O. Scully[1,3,4#]

[1]*Texas A&M University, College Station, TX 77843, USA*
[2]*Xi'an Jiaotong University, Xi'an, Shaanxi 710049, China*
[3]*Baylor University, Waco, TX 76798, USA*
[4]*Princeton University, Princeton, New Jersey 08544, USA*

Two-dimensional (2D) materials beyond graphene such as transition metal dichalcogenides (TMDs) have unique mechanical, optical and electronic properties with promising applications in flexible devices, catalysis and sensing. Optical imaging of TMDs using photoluminescence and Raman spectroscopy can reveal the effects of structure, strain, doping, defects, edge states, grain boundaries and surface functionalization. However, Raman signals are inherently weak and so far have been limited in spatial resolution in TMDs to a few hundred nanometres which is much larger than the intrinsic scale of these effects. Here we overcome the diffraction limit by using resonant tip-enhanced Raman scattering (TERS) of few-layer MoS$_2$, and obtain nanoscale optical images with ~ 20 nm spatial resolution. This becomes possible due to electric field enhancement in an optimized subnanometre-gap resonant tip-substrate configuration. We investigate the limits of signal enhancement by varying the tip-sample gap with sub-Angstrom precision and observe a quantum quenching behavior, as well as a Schottky-Ohmic transition, for subnanometre gaps, which enable surface mapping based on this new contrast mechanism. This quantum regime of plasmonic gap-mode enhancement with a few nanometre thick MoS$_2$ junction may be used for designing new quantum optoelectronic devices and sensors.

Molybdenum disulfide (MoS$_2$) is a layered 2D TMD material with a graphene-like hexagonal arrangement of Mo and S atoms covalently bonded into single layer S-Mo-S units (Fig. 1b) which are stacked and held together by van der Waals (vdW) interactions.[1,2] The unique mechanical, optical and electronic properties of the monolayer (ML) and few-layer (FL) MoS$_2$ are attractive for photovoltaic, optoelectronic and sensing applications [3-8]. The device performance strongly depends on the quality of materials and contact interfaces which varies on the nanoscale. Therefore, to properly understand and control these materials and devices, it is necessary to study their properties with nanometre spatial resolution. While atomic force microscopy (AFM)[9,10] and scanning tunneling microscopy (STM) [11-13] provide rich information about mechanical and electronic properties of TMDs, optical spectroscopic techniques such as photoluminescence (PL) [5,6,14,15] and Raman scattering [16-23] provide important complementary optical information about structure and dynamics. Previous optical studies of MoS$_2$ grown by chemical vapor deposition (CVD) revealed interesting relaxation properties of edge states and grain boundaries using near-field scanning optical microscopy with > 60 nm resolution [24]. Their PL signals showed ~ 300 nm wide edge regions and more than 80 nm wide disordered grain boundaries. These results were contrary to the expected ~ 20 nm atomically-thin termination edge states and were attributed to disorder. Further investigation of these effects requires increasing the spatial resolution and can be performed using other optical techniques such as Raman spectroscopy.

Tip-enhanced Raman scattering (TERS) is a vibrational spectroscopic technique with nanoscale spatial resolution enabled by the local field enhancement of the incident laser field at the nanosize apex of a plasmonic tip[25-30]. A major challenge of TERS is to overcome the background signal from the diffraction-limited laser focus. It has been overcome in several cases by placing the sample between the plasmonic substrate and the tip in the so-called "gap-mode" configuration[31-33]. Near-field enhancement and the corresponding TERS signals are classically expected to increase with the decrease of the gap between the tip and the sample[25,34-36], similar to the case of nearly-touching nanoparticles [37-41]. Therefore, decreasing the tip-sample gap is expected to improve the TERS imaging contrast. Quantum-mechanically, however, there is a limit to the maximum field enhancement for quantum emitters placed in subnanometre gaps leading to the quantum plasmonics regime [38,39,42-48]. Gap-dependent plasmonic phenomena were predicted and observed in nanogap junctions between metal surfaces which required the quantum mechanical description to explain the saturation and quenching behavior of the near-field enhancement [38,39,42,45,49]. This quantum limit poses restrictions on surface-plasmon enhanced optical signals and needs to be better understood and controlled. Diffraction-limited Raman imaging was previously used to investigate layer-dependence[23], electron-phonon coupling [20], and quality of vdW interfaces [21] in FL-$MoS_2$ and heterostructures. Higher spatial resolution would help better resolving the physical processes at the edges and in heterostructure contacts. However, Raman scattering is inherently weak and has not yet been observed in TMDs at the nanoscale using any subdiffraction optical imaging technique. TERS from other low-dimensional materials such as graphene [50-53] and carbon nanotubes (CNTs) [54-57] was previously observed. However, graphene and CNTs have larger Raman cross-sections then most TMDs and are more favorable for TERS imaging.

Here, we use subnanometre-gap resonant TERS to obtain first nanoscale Raman images of FL-$MoS_2$ flakes on the gold substrate. Such gap-mode configuration is expected to show classical enhancement where smaller gaps lead to larger optical signals. Surprisingly, our results show an opposite behavior revealing several enhancement regimes including non-classical signals decreasing for smaller gaps. We demonstrate quantum plasmonic coupling between gold and FL-$MoS_2$ which could offer unique advantages, for example, in flexible sensors with new optical properties. Moreover, since the quantum quenching behavior depends on the material probed, it provides enhanced contrast for surface and edge sensing and mapping.

**Results**

**Tip-enhanced Raman scattering (TERS) of FL-MoS₂.** The schematic of the experimental setup is shown in Fig. 1a. We used a gold coated tip with ~ 20 nm apex radius to enhance Raman signals from exfoliated FL-MoS₂ flakes deposited on the atomically-flat gold substrate (Fig. 1c). We performed resonant TERS measurements by exciting FL-MoS₂ into the A-exciton band using a 660 nm p-polarized laser (for details see Supplementary Fig. 1). Fig. 1d shows the measured optical signals from the location marked with a white dot in Fig. 1c as a function of the tip-sample distance in the units of the sample z-axis displacement (see Fig. 2 for conversion from the sample z-axis displacement to the tip-sample distance). We varied the tip-sample distance from a few nanometres down to a few Angstroms to achieve the maximum signal enhancement. The TERS measurement location marked by the white dot in Fig. 1c corresponds to the MoS₂ thickness of four layers as determined by atomic force microscopy (AFM). At ~ 4 nm z-axis displacement, the signal intensity suddenly increases leading to the so-called "snap to contact" (see below) [44]. At that moment, the actual distance between the tip and the sample is ~ 3 nm. The thickness of the four-layer MoS₂ is ~ 2.7 nm. Therefore, the distance between the tip and the gold substrate is ~ 5.7 nm, which gives rise to the plasmonic gap mode which is crucial for signal enhancement due to formation of the electric field hot spots in the tip-substrate gap.[31-33] No tip enhancement was observed in similar experiments using the gold tip from FL-MoS₂ deposited on a glass substrate. Fig. 1e shows the spectra before (~ 5.8 nm, red) and after (~ 0.33 nm, green) the contact between the tip and the FL-MoS₂. The observed transitions are assigned to the two first-order Raman-active vibrational modes at the $\Gamma$ point of the Brillouin zone (BZ): in-plane $E_{2g}^1$ and out-of-plane $A_{1g}$ at 382 cm⁻¹ and 408 cm⁻¹, respectively.[58] The mode at 177 cm⁻¹ is the difference combination mode of the out-of-plane and longitudinal acoustic phonon at the M point: $A_{1g}(M) - LA(M)$.[22,58] The spectra in the range between 450 cm⁻¹ and 490 cm⁻¹ show a double mode feature. The most pronounced mode at 464 cm⁻¹ is the $A_{2u}$ mode, while the mode on its shoulder is associated with the $LA(M)$ phonon.[58] In the spectral range higher than 490 cm$^{-1}$, there are four second-order modes at 526 cm$^{-1}$ ($E_{1g}(M) + LA(M)$), 570 cm$^{-1}$ ($2E_{1g}$), 599 cm$^{-1}$ ($E_{2g}^1 + LA(M)$) and 642 cm$^{-1}$ ($A_{1g}(M) + LA(M)$) [58]. These different modes were enhanced by different factors due to their different polarization selection rules. When the tip approaches the sample it causes depolarization of the incident p-polarized laser field which leads to the enhancement of, for example, s-polarized in-plane modes. The estimated enhancement factors (EFs) of the strongest modes are given in Supplementary Fig. 7 and Supplementary Table 1. The in-plane $E_{2g}^1$ mode shows the largest enhancement due to the depolarization of the incident laser field by the tip. The dashed lines in Fig. 1e show the fittings of the photoluminescence (PL) background signals.

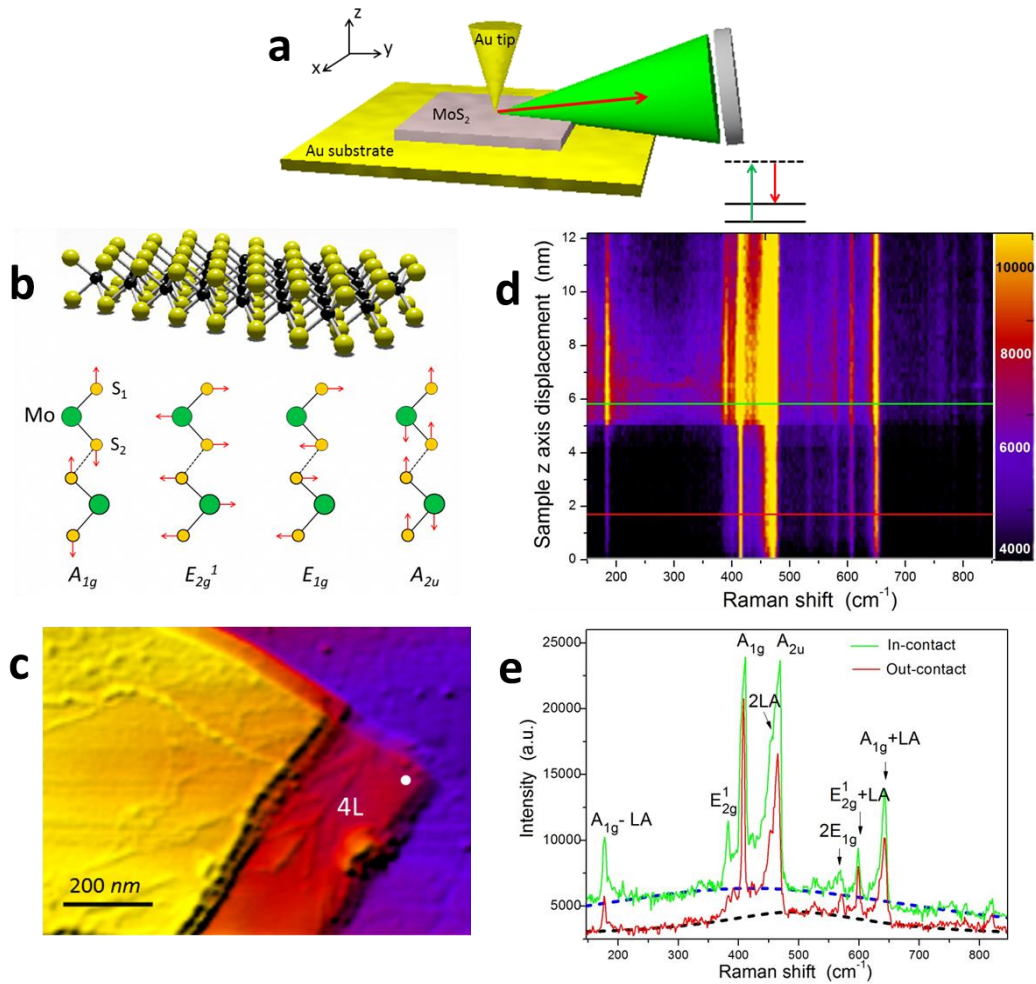

**Figure 1**. **Resonant tip-enhanced Raman scattering (TERS) of FL-MoS$_2$. a**. Schematic of the experimental setup: 660 nm laser (green cone) is focused on the gold tip in contact with the FL-MoS$_2$ flake on the gold substrate. The emitted light (red arrow) is detected in the backscattering geometry. Inset shows the Raman process. **b.** Structure of the monolayer MoS$_2$ and four observed vibrational modes of the FL-MoS$_2$. **c**. AFM image of the FL-MoS$_2$ flake. White dot marks the location from which the tip-enhanced optical signals in **d** and **e** were obtained. **d.** Sample z-axis displacement-dependence of the optical signal from the location marked by the white dot in **c**. **e.** Optical spectra for the tip-sample distance of 5.8 nm (red line, out-of-contact) and 0.33 nm (green line, in-contact) which correspond to the red and green lines in **d**, respectively. Dashed lines show the fittings of the photoluminescence background signals. The observed Raman transitions are labeled in **e** according to the vibrational modes shown in **b**.

**Subnanometre-gap dependence with sub-Angstrom control.** Resonant excitation of surface plasmons by a metallic tip near the metallic surface gives rise to a large electric field enhancement described by the image charge model [59]. We investigate the dependence of the signal enhancement on the tip-sample distance by approaching the gold tip to gold, FL-MoS$_2$, and carbon nanotube (CNT) samples with sub-Angstrom precision. This is achieved by decreasing the tip-sample distance using contact-mode AFM force-distance measurements (Fig. 2). Fig. 2a shows the applied cantilever force as a function of the sample z axis (piezo) displacement during the tip-sample approach. At ~ 4 nm displacement, the gradient of the tip-sample force exceeds the elastic constant of the cantilever and the tip jumps onto the sample surface in the so-called "snap to contact".[60] Afterwards, the movements of the tip and the sample are linearly coupled which is reflected by the straight "contact line." We denote the starting point of the contact line as "vdW contact" as the onset of the localized repulsive force which manifests that the tip-sample distance reached the vdW diameter (~ 0.3 nm). After the vdW contact, the total force on the sample remains attractive for a while as adsorbed molecules, such as water vapor under ambient conditions, contribute to the adhesive force. Fig. 2b shows the displacement-repulsive force curve after the vdW contact resulting from the subtraction of the adhesive force from the total applied force. The spike at ~ 10.7 nm displacement on the curve with ~30 nN repulsive force indicates that the tip penetrated the MoS$_2$ flake.[61] Therefore, for the tip-sample distance dependence analysis below we only consider the data before this breaking point. We obtain the tip-sample distance control before the vdW contact from the linear relation between the applied force and the cantilever deflection with the accuracy of ~ 0.16 nm (see Supplementary Fig. 3). However, after the vdW contact, we use the repulsive force from the Lennard-Jones model ($f_{rep} = Ad^{-13}$) to convert the repulsive force into the tip-sample distance with sub-Angstrom accuracy of ~ 0.0014 nm (see Supplementary Fig. 4). We determine value of A = 2.2*10$^{-7}$ nN nm$^{13}$ at the breaking point when the distance between Au and S atoms is close to the sum of their radii ~ 2.42 A$^o$.[62] Fig. 2c shows the corresponding conversion of the repulsive force into distance. More details are given in Supplementary Fig. 2 and Supplementary Note 1.

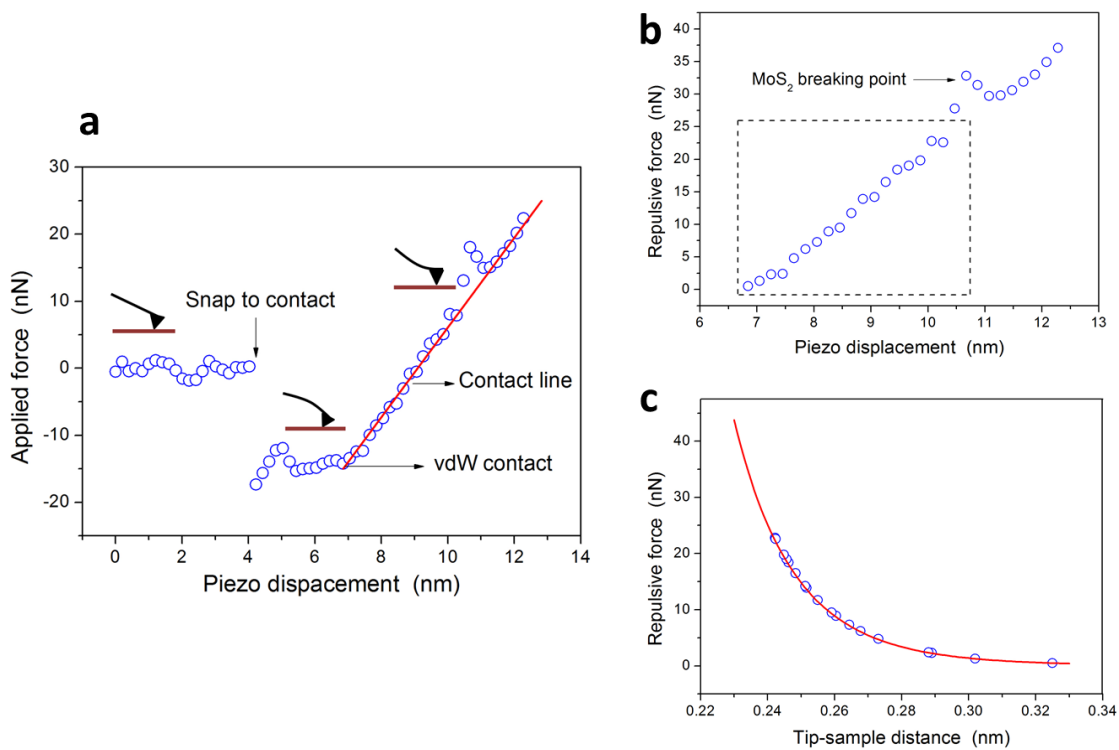

**Figure 2**. **Subnanometre-gap distance dependence of FL-MoS$_2$-gold tip interaction with sub-Angstrom control. a.** Force-displacement curve measured by AFM (circles) and linear fit (red line) of the contact line. Three tip-sample interaction moments are depicted: (i) snap to contact; (ii) vdW contact; (iii) contact line. **b.** Repulsive force after vdW contact. At the MoS$_2$ breaking point, the distance between Au and S atoms is ~ 0.24 nm. Dashed rectangle selects the linear range of data used for converting repulsive force into subnanometre tip-sample distance via Lennard-Jones model (red line) in **c**. This procedure provides sub-Angstrom control of the tip-sample distance and was used to observe quantum quenching and Schottky-Ohmic transition in FL-MoS$_2$.

Fig. 3a shows the measured PL intensity of the gold tip in the vicinity of the atomically-flat gold substrate at 1.82 eV (~ 510 cm$^{-1}$) increasing with the decrease of the tip-sample distance before the vdW contact as predicted by the classical model. At ~ 2 nm separation, the PL intensity steeply increases, and reaches the maximum when the tip contacts the sample at ~ 0.33 nm corresponding to the typical Au vdW diameter. Afterwards, the cantilever deflection increases linearly with the decrease of the tip-sample distance, due to the increase of the repulsive force between the tip and the sample (see Fig. 2 and Supplementary Fig. 2), accompanied by the rapid decrease of the PL as predicted by the quantum model. Similar PL quenching of the gold tip near a gold substrate was previously observed using a combination of AFM and scanning tunneling microscopy (STM) and was attributed to be due to quantum tunneling and nonlocal effects.[48] However, our experiments are performed using only AFM without bias which could explain the differences in the onset of the PL quenching and in the distance dependence of the Raman signals. We used the same tip to repeat the distance-dependence measurements several times to confirm reproducibility. The reproducible results shown in Supplementary Fig. 5 prove that the tip was not broken during the measurements. As the tip-sample distance decreases to the vdW diameter, an effective electron wave function overlap between the tip and the substrate is established, forming a conductive tunneling junction. The capacitive coupling before the contact is reduced by the conduction current between the tip and the substrate leading to charge transfer plasmons.[37,63,64] The classical coupling of the tip and substrate via image charges results in the large near field enhancement in few nanometre gaps.[31] The conduction current in the gap suppresses the charge accumulation on the surfaces, decreasing the near fields and quenching the PL signals at subnanometre gaps in the quantum coupling regime. These classical and quantum coupling schemes are shown for the gold tip coupled to the gold substrate in Fig. 3d.

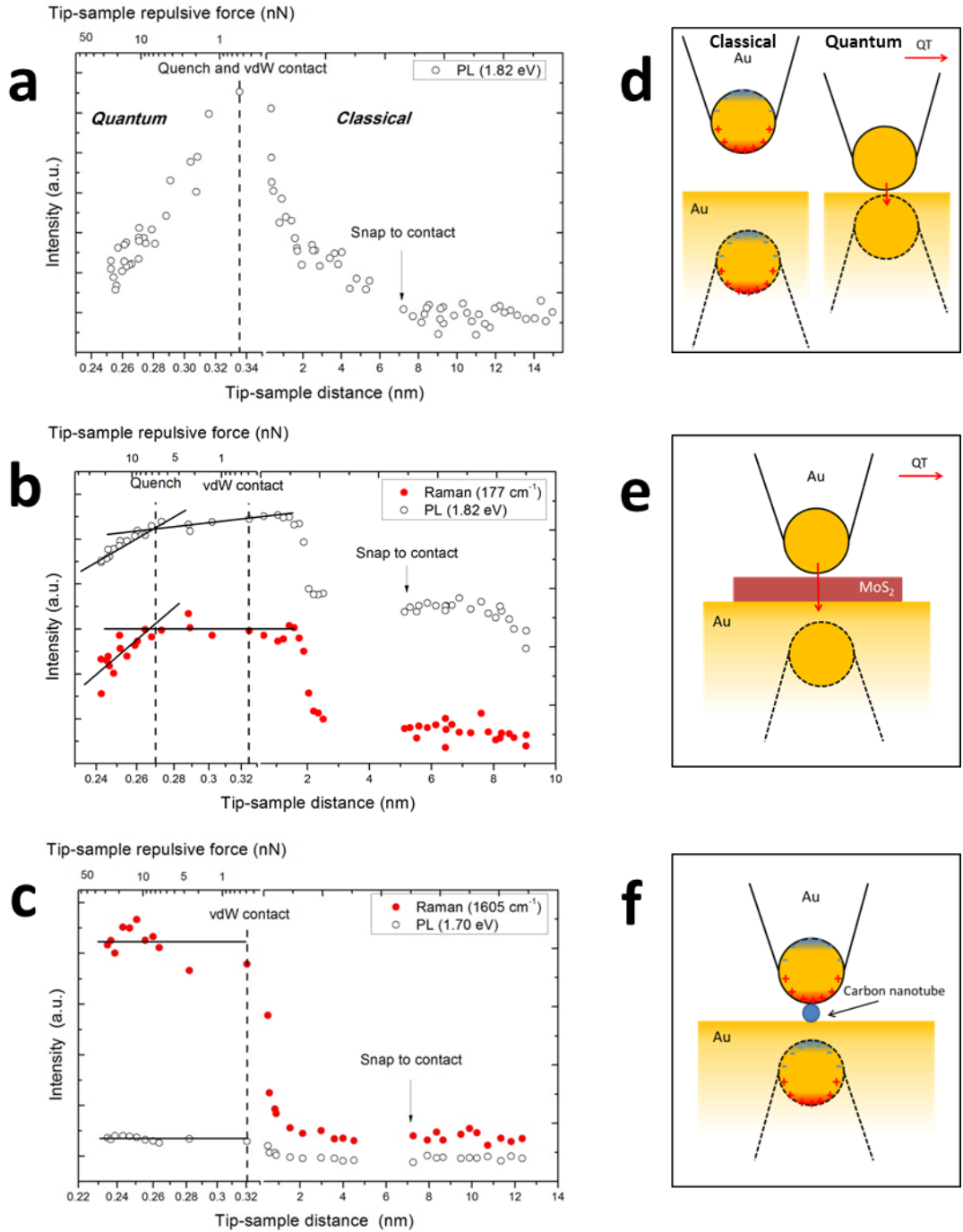

**Figure 3. Subnanometre-gap transition between classical and quantum regimes.** Tip-sample distance-dependence of the tip-enhanced optical signals from the gold tip near the flat gold substrate without **a**, and with the FL-MoS$_2$ **b** and a carbon nanotube **c** junctions. Photoluminescence (PL) and Raman signals are shown as open and red filled circles, respectively. Vertical dashed lines denoted by "Quench" and "vdW Contact" show the moments at which the signals begin to decrease and the tip-sample

distance approaches the van der Waals (vdW) diameter, respectively. Classical and quantum coupling schemes of the gold tip **d**, FL-MoS$_2$ **e** and carbon nanotube **f** on gold substrates. Red arrows represent quantum tunneling (QT) currents. Dashed lines show tip images in the substrates.

Quantum plasmonic regime of the near field and optical signal enhancement in coupled metallic nanostructures has been well explored.[47,49] Here we investigate the corresponding quantum limit in the hybrid system consisting of the FL-MoS$_2$ between two gold surfaces. Fig. 3b shows the tip-sample distance-dependence of the FL-MoS$_2$ sample PL at 1.82 eV and Raman signal at 177 cm$^{-1}$. Both signals increase with the decrease of the tip-sample distance before the vdW contact, as in the case of the gold substrate PL. However, after the vdW contact the distance-dependences for the FL-MoS$_2$ and gold substrates are different. There is a sharp turnover point for the PL from gold at the moment of vdW contact, whereas the FL-MoS$_2$ signals stay nearly constant and start decreasing for the tip-sample distances smaller than the vdW diameter. The quench and vdW contact occur at the gap sizes of ~ 0.27 and 0.32 nm, respectively, for FL-MoS$_2$, while they both occur at the same gap size of ~ 0.33 nm for gold.

Fig. 3c shows a control measurement of the distance-dependent PL and Raman signals for the G band of the carbon nanotube (CNT) on the gold substrate. There is no significant quenching after the vdW contact. Fig. 3f indicates that the ~ 1.2 nm CNT diameter is smaller than the tip apex curvature. Therefore, the bonding charges are transferred only partially, which corresponds to the screened plasmon modes and a small decrease of the field enhancement. However, the CNT-gold junction forms a Schottky contact with ~ 0.3 eV Schottky barrier height (SBH) and suppressed electron injection.[65] The comparison between FL-MoS$_2$ and CNT shows that only certain semiconductor materials with good conductivity and Ohmic contacts with gold can exhibit quantum plasmonic coupling.

**Subnanometre-gap nanoimaging**. Fig. 4 shows subnanometre-gap tip-enhanced nanoimaging of another FL-MoS$_2$ flake. A region highlighted by the black rectangle in the AFM image in Fig. 4a was chosen for spatial mapping of the tip-induced optical signals in Figs. 4b and 4c. The force applied by the tip on the sample was set to a value corresponding to the tip-sample distance of ~ 0.25 nm which corresponds to the quantum regime. The weak FL-MoS$_2$ PL signal at ~1.82 eV (510 cm$^{-1}$) overlaps with the strong PL emission of the gold tip.[66,67] Figs. 4b and 4c show the comparison of the PL and Raman mapping. Surprisingly, the PL map in Fig. 4b shows a high contrast with the enhanced PL signal at the edge of the FL-MoS$_2$ flake and quenching on the gold substrate and on the inner area of the flake. These effects are attributed to the subnanometre-gap quantum regime. Quenching of the PL emission of the gold tip on the gold substrate is stronger than on the FL-MoS$_2$, resulting in a high imaging contrast. Fig. 4d shows the AFM (solid line), PL (open circles) and Raman (red filled circles) profiles across the edge of the MoS$_2$ flake which correspond to the yellow dashed line in Fig. 4a. The maxima of the PL and Raman profiles occur at the edge of the flake, where only a part of the tip is in contact. Fig. 4e shows the corresponding model of the partially screened subnanometre gap plasmon imaging. Only the charges in the vicinity of the gap flow through the junction reducing the quenching effect. Therefore, the tip-induced signals are quenched both on the gold substrate and on the inner part of the flake but the quenching is suppressed at the edge. The thickness of the FL-MoS$_2$ in the highlighted black rectangle is ~ 2.8 nm which corresponds to four layers. The size of the observed FL-MoS$_2$ optical edge profile is ~ 20 nm which corresponds to the size of the tip but is significantly smaller than the size of the excitation laser spot (~1 μm). This edge profile was used to estimate the TERS spatial resolution of ~ 20 nm (see Supplementary Fig. 6).

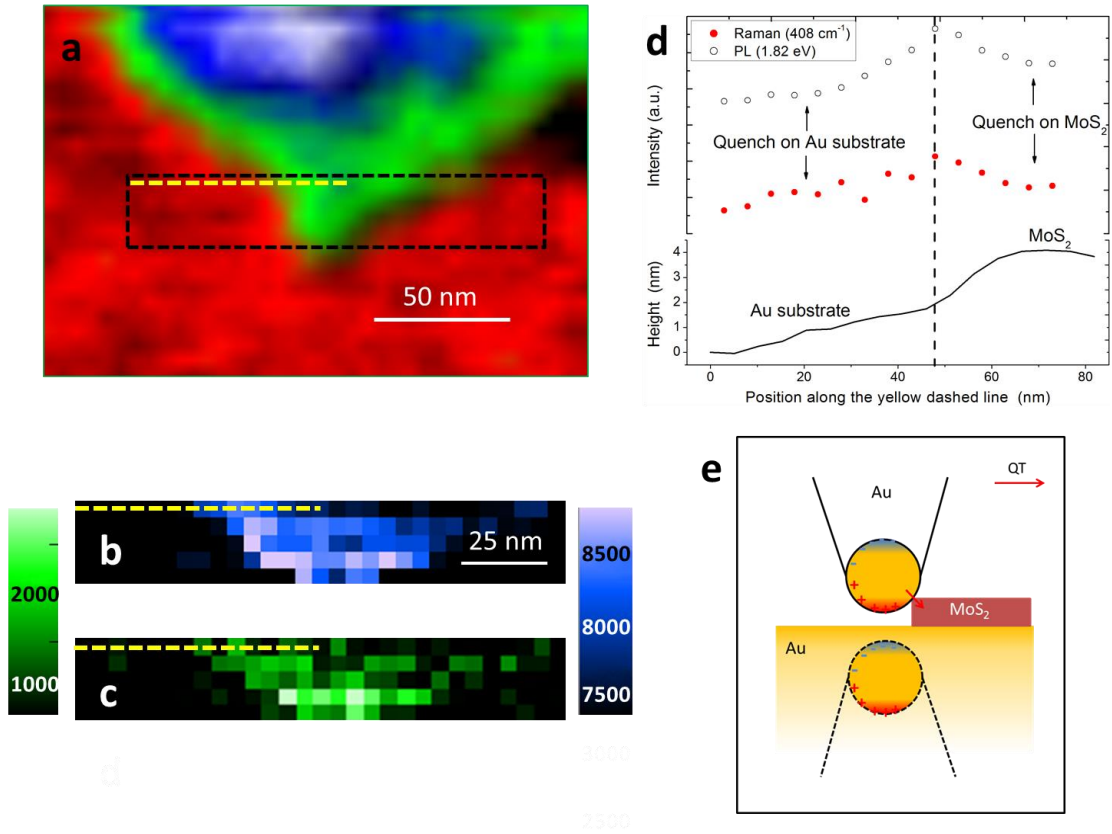

**Figure 4. Subnanometre-gap tip-enhanced nanoimaging**. **a.** AFM image of the FL-MoS$_2$ flake. Tip-enhanced optical images corresponding to the black dashed rectangle in **a** for photoluminescence (PL) at 1.82 eV (510 cm$^{-1}$) **b**, and Raman transition at ~ 408 cm$^{-1}$ **c**. **d** PL (open circles), Raman (red filled circles) and AFM (solid line) line profiles which correspond to the yellow dashed line in **a**. **e.** Partially screened subnanometre-gap coupling scheme for the edge of the FL-MoS$_2$.

## Discussion

Several mechanisms can explain the decrease of the tip-enhanced optical signals at subnanometre gaps. Classical resonance energy transfer and quantum tunneling (QT) effects quench PL at large and small gaps, respectively.[48] However, both mechanisms have no direct influence on the Raman scattering process due to its nearly instantaneous relaxation. Most commonly, the quench of the Raman signals is due to the decrease of the electric fields in the gap which strongly relies on the capacitive coupling between the tip and the substrate. Quantum tunneling between the gold tip and the FL-MoS$_2$ can reduce the charge accumulation on the surface of the tip and decrease the corresponding electric fields.[68] Figs. 3d – 3f show the coupling schemes in the three investigated sample configurations. The FL-MoS$_2$ – gold system is described as a metallic sphere-plane configuration with the FL-MoS$_2$ conductive junction, whose conductivity impacts the optical response of the system.[69] For high conductivity and large aspect ratio, charge transfer plasmons can be generated. For partial coverage of the junction, the charge transfer takes place in the vicinity of the junction, giving rise to screened plasmons with partial charges at both metallic surfaces.[69] This requires an abundant number of carriers on both sides of the junction, which may be increased due to photo-excitation. Previous work showed that MoS$_2$ photocurrent can be enhanced by surface plasmons of gold nanoparticles.[70] However, for metal-semiconductor junctions, the SBH also significantly impacts the conductive electron injection. The SBH of the FL-MoS$_2$ – gold junction is low and favorable for the electron injection (~ 50 meV).[10,71] However, formation of the Ohmic contact is sensitive to the distance between the metal and MoS$_2$. The hybridization and overlap of the electron wave functions between the Au and Mo atoms play a dominant role in the electron injection from the gold tip to MoS$_2$.[62] The distance between the Mo and S atomic layers in the monolayer MoS$_2$ is ~ 0.24 nm. When the gold tip contacts the S atomic layer, the distance between the Au atoms at the tip apex and the Mo atoms beneath the first S atomic layer is larger than the vdW diameter. The corresponding SBH is larger than 50 meV preventing the formation of the Ohmic contact. However, even though the tip gets closer to the MoS$_2$ surface by only 0.06 nm which still cannot make a good contact between the Au and Mo atoms, an effective electrical contact can be formed between the gold tip and the MoS$_2$ sample due to the formation of the gap state, as S atoms mediate the hybridization between the Au and Mo atoms.[72] This mechanism can explain why the quench of the FL-MoS$_2$ tip-enhanced signals and the vdW contact do not take place simultaneously. In contrast, when the gold tip gets in contact with the gold surface at the vdW diameter separation, the effective overlap of the electron wave functions between the tip and the substrate gives rise to the effective electron injection. For the FL-MoS$_2$ junction, the formation of the gap state requires a shorter distance than the vdW diameter.[72] Fig. 3b provides evidence for the formation of an Ohmic contact in the FL-$MoS_2$ – gold junction when a stronger force is applied at subnanometre gaps. This makes $MoS_2$ an attractive material for conductive junctions.

In summary, we demonstrated the first tip-enhanced Raman scattering signals of FL-$MoS_2$ on a gold substrate, which revealed the quantum coupling regime. We showed a new approach for studying the nanooptical properties of gold-$MoS_2$ junctions via tip-sample distance dependence of subnanometre-gap Raman and PL signals. This allowed for distinguishing ~ 20 nm edge states and nanoscale inhomogeneous structural features in FL-$MoS_2$. The observed quantum quenching behavior depends on the material probed and provides enhanced contrast for surface sensing and mapping. This quantum imaging contrast can be used for future studies of grain boundaries and defects in 2D materials. Quantum coupling can also be used to enhance the sensitivity of single molecule imaging when the substrate is non-metallic. Subnanometre-gap Ohmic contacts and quantum plasmonic effects may be used to improve the performance of $MoS_2$-based devices such as Schottky diodes and phototransistors. Other techniques can be used to further enhance the near fields and Raman signals based on coherence [73] and gain [74]. Our work opens new possibilities to explore quantum effects in plasmonic chemical sensors and photo-devices based on 2D materials.

**Methods**

**Materials and sample preparation.** Tips were gold coated (AIST NT) with ~ 20 nm apex radius. FL-$MoS_2$ flakes were prepared using mechanical exfoliation from molybdenite (SPI), and were deposited on the atomically-flat Au/$SiO_2$ substrate (Platipus). Single-wall carbon nanotubes (~ 70% metallic) on atomically-flat gold substrate were provided by AIST-NT.

**Tip-enhanced Raman scattering (TERS).** All atomic force microscopy (AFM) and TERS measurements were performed using a combined scanning probe microscopy (SPM) system (OmegaScope-R, AIST-NT) and a Raman microscope (LabRAM HR4000, Horiba). Tapping-mode AFM was used for measuring sample topography. Contact-mode AFM was used for TERS distance-dependence and imaging measurements. A cw diode laser (659.38 nm, 1.88 eV) was focused onto the gold tip apex with an objective lens (NA = 0.9) in the side illumination, polarized along the tip axis. The signal was collected in the backscattering configuration with a long-pass 661.56 nm edge filter, and was detected with a grating spectrometer coupled to a CCD camera. Reproducibility of TERS measurements was checked on several samples (see Supplementary Fig. 5)

**Acknowledgements**

We acknowledge the support of the National Science Foundation Grants No. EEC-0540832 (MIRTHE ERC), No. PHY-1068554, No. PHY-1241032 (INSPIRE CREATIV), and No. PHY-1307153, the Office of Naval Research, and the Robert A. Welch Foundation (Awards A-1261 and A-1547).

**Author Contributions**

DVV, AVS, ZZ and MOS conceived the idea and designed the experiments. MH prepared the sample. YZ, DVV, SQ and AMS performed the experiments. YZ, DVV and AMS analyzed the data. YZ, DVV, AMS, AVS, ZZ and MOS wrote the paper. *These authors contributed equally to this work.

**Additional Information**

**Supplementary Information** accompanies this paper.

Supplementary Information for

# Quantum limit in subnanometer-gap tip-enhanced nanoimaging of few-layer MoS$_2$

Yingchao Zhang[1,2], Dmitri V. Voronine[1,3*], Shangran Qiu[1,2], Alexander M. Sinyukov[1], Mary Hamilton[3], Alexei V. Sokolov[1,3], Zhenrong Zhang[3] and Marlan O. Scully[1,3,4]

[1]*Texas A&M University, College Station, TX 77843, USA*
[2]*Xi'an Jiaotong University, Xi'an, Shaanxi 710049, China*
[3]*Baylor University, Waco, TX 76798, USA*
[4]*Princeton University, Princeton, New Jersey 08544, USA*

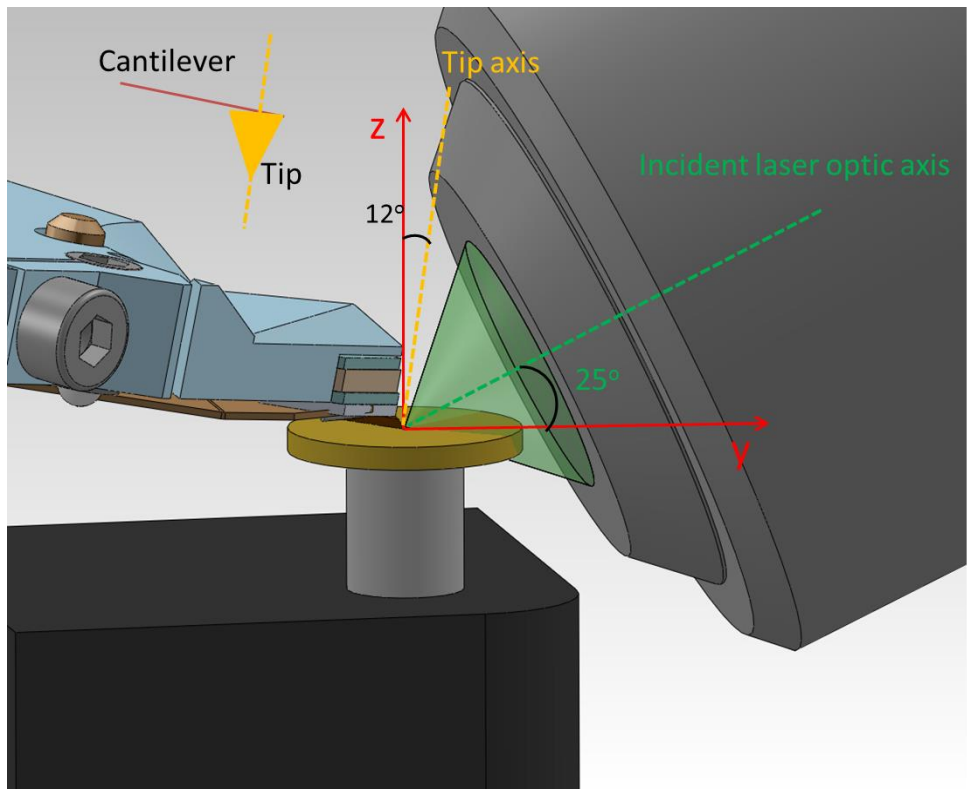

**Supplementary Figure 1. Schematic of the tip-enhanced Raman scattering (TERS) experiments using a combination of scanning probe microscope (OmegaScope-R, AIST-NT) and Raman microscope (LabRAM HR Evolution, Horiba)**. 660 nm laser (green) is focused on the gold tip. The angle between the incident laser optical axis and y axis is 25º. The angle between the tip axis and z axis is 12º. The frame of axes is the same as shown in Fig.1a. The scattered Raman signal is collected by the illumination objective. The tip and the laser are stationary, while the sample is scanned in x, y, and z-directions.

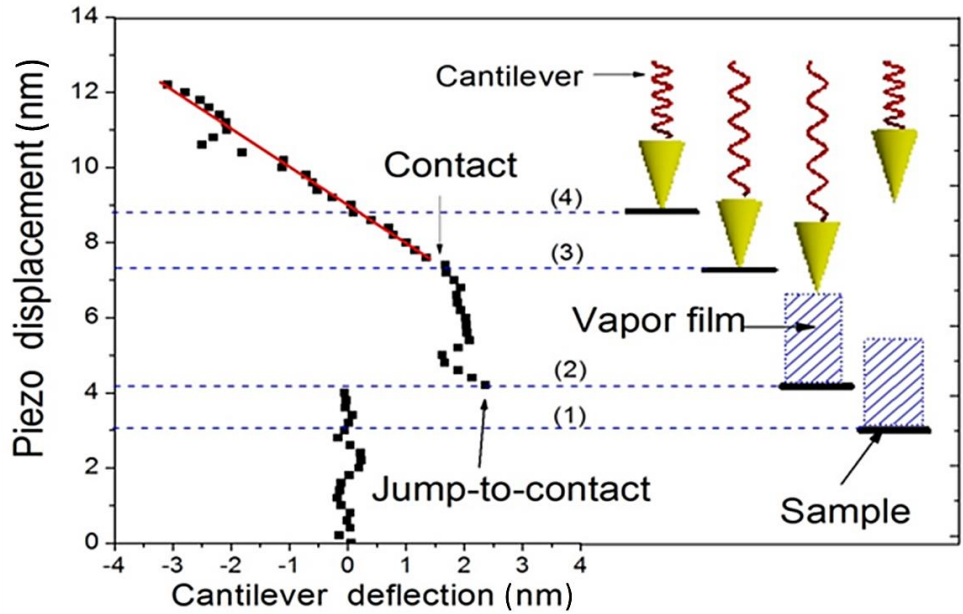

**Supplementary Figure 2. Displacement-deflection curve**. The cantilever deflection is plotted vs the tip-sample displacement of the AFM piezo actuator. The red solid straight line shows that after the tip-sample contact the cantilever is deflected linearly, as the tip and the sample move upwards together. The corresponding tip and sample states at four typical moments are highlighted by dashed lines.

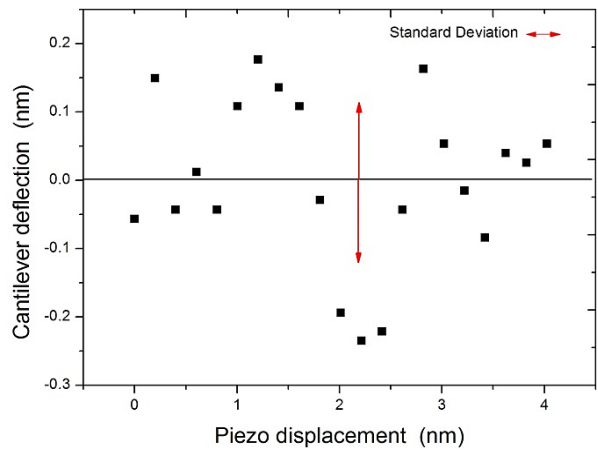

**Supplementary Figure 3. Deviation of tip-sample distance before vdW contact.** Before vdW contact, the tip-sample distance deviation results from both cantilever and piezo displacements. Fluctuation of the cantilever deflection before vdW contact has 0.12 nm standard deviation. The piezo displacement has the inherent deviation ~ 0.1 nm. Thus, the total deviation of the tip-sample distance before vdW contact is $\sqrt{0.12^2 + 0.1^2} \approx 0.16$ nm.

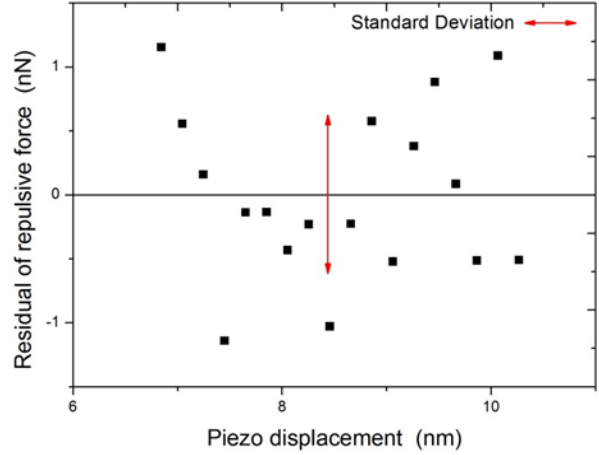

**Supplementary Figure 4. Deviation of tip-sample distance after vdW contact.** After vdW contact, the deviation of the tip-sample distance is mainly due to the fluctuation of the repulsive force which is obtained by the residual of the linear contact line fitting in Fig. 2b. The deviation of the repulsive force is 0.67 nN. According to the error transfer relation $\left|\frac{\Delta f}{\bar{f}}\right| = 13 \left|\frac{\Delta d}{\bar{d}}\right|$, where $f_{rep} = Ad^{-13}$, and $\bar{f}$ is the mean repulsive force (11.05 nN), and $\bar{d}$ is the mean distance after vdW contact (0.3 nm), the deviation of the tip-sample distance after vdW contact is 0.0014 nm.

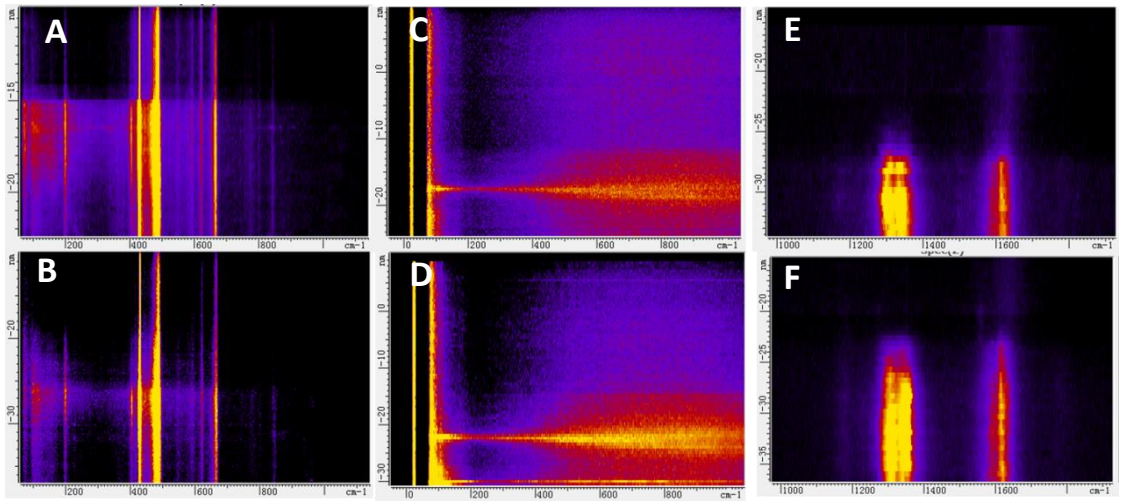

**Supplementary Figure 5. Reproducibility of TERS measurements.** Similar results were obtained from several spots in FL-MoS$_2$ flake (**a** and **b**), on gold substrate (**c** and **d**) and on carbon nanotubes (**e** and **f**).

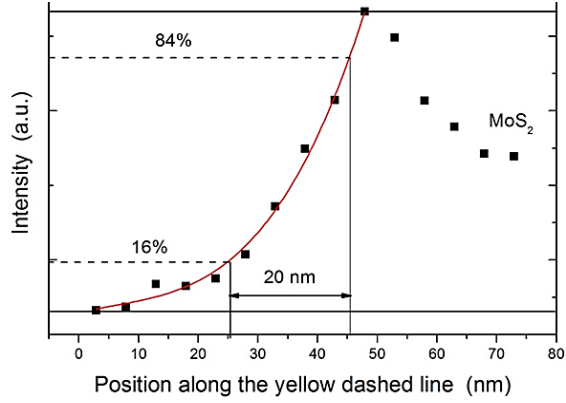

**Supplementary Figure 6. Spatial resolution of optical imaging.** Optical profile along the yellow dashed line in Fig. 4a provides an upper bound for spatial resolution of ~ 20 nm for optical imaging.

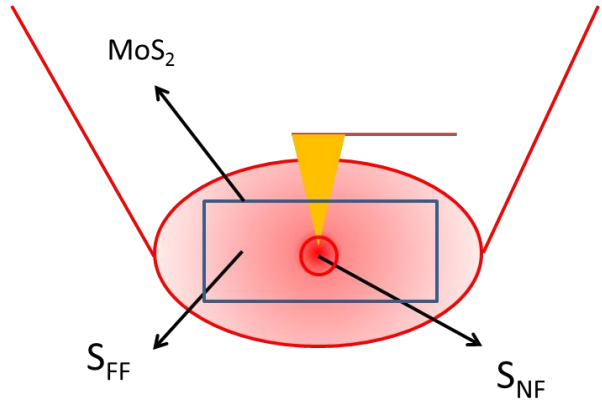

**Supplementary Figure 7.** Schematic of the incident and local fields for enhancement factor (EF) calculations. Large red ellipse represents the focused incident laser electric field which covers the FL-MoS$_2$ flake shown by the blue rectangle. The area of the flake was estimated from the AFM image in Fig. 1c to be $S_{FF}$ = 200*800 nm = 160000 nm$^2$ and was used as the far-field normalization factor in the EF equation [8]:

$$EF = \left(\frac{FF + NF}{FF} - 1\right)\frac{S_{FF}}{S_{NF}}$$

Here, the numerator FF + NF and the denominator FF are the intensities of the optical signals with the gold tip in contact and out of contact with the FL-MoS$_2$, respectively. The near field normalization factor $S_{NF} = \pi(R_{NF})^2 = 1256$ nm$^2$ was estimated by assuming the area of the near field spot to be approximately equal to the size of the tip with the radius $R_{NF}$ = 20 nm. The calculated EFs for selected strongest Raman transitions of the FL-MoS$_2$ flake in Fig. 1 with 0.32 nm gap are shown in Supplementary Table 1.

**Supplementary Table 1.** Enhancement factors (EFs) for selected strongest Raman transitions of the FL-MoS$_2$ flake in Fig. 1 with 0.32 nm gap estimated as shown in Supplementary Fig. 7.

| Transition (cm$^{-1}$) | EF |
|---|---|
| A$_{1g}$-LA (177) | 124 |
| E$_{2g}^1$ (382) | 284 |
| A$_{1g}$ (408) | 3 |
| A$_{2u}$ (464) | 58 |
| E$_{2g}^1$+LA (599) | -6 |
| A$_{1g}$+LA (642) | 46 |

## Supplementary Note 1
## Tip-sample distance dependence analysis

The AFM cantilever can be regarded as a spring with the force - deflection dependence given by Hooke's law: $F = k\delta_c$, where $\delta_c$ is the cantilever deflection. The spring constant of the cantilever used in this work was $k = 2.8 N/m$. During the measurements, the tip was kept stationary and the sample moved upwards in discrete steps (0.2 nm), driven by the piezoelectric actuator with the precision of ~ 0.1 nm. The tip-sample piezo displacement and the corresponding cantilever deflection were recorded and used to calculate the actual tip-sample distance by following the procedure summarized in Fig. 2 and described in more detail below. Supplementary Fig. 2 shows the plot of the cantilever deflection versus the tip-sample piezo displacement. We denote four typical moments to illustrate the behavior of the tip during the tip-sample approach. During moment (1), the sample is far from the tip, and the interaction force between the tip and the sample is negligible. Therefore, the cantilever deflection stays constant. At moment (2), the displacement-deflection curve reveals a singularity, which means that the cantilever is dragged and "jumps" onto the sample. This is the so-called "Snap to contact".[1] In the high vacuum and low temperature conditions, the "Snap to contact" occurs typically at the subnanometer length scale of tip-sample separation, when the derivative of the force versus tip-sample separation exceeds the spring constant of the cantilever. However, in ambient environment, a liquid film (vapor) can be adsorbed on the sample surface.[2, 3] The enhanced adhesive force of the vapor film gives rise to a sudden "jump" over ~ 2 nm, when the tip is at the position of ~ 5 nm above the sample surface. After the contact, the tip penetrates the film which can only exert a limited repulsive force, while the attractive force is still dominant. Therefore, the cantilever deflection slightly decreases.[4] At moment (3), the tip breaks through the film and contacts the sample surface. The red line in Supplementary Fig. 2 denotes the so-called "contact line," which shows the linear coupling of the sample and the tip movements.[3, 5] We assumed that at the vdW contact the repulsive force is 0.5 nN, which corresponds to 3.26 A° vdW diameter.

# Supplementary References